\documentclass[10pt,oneside,english]{amsart}
\usepackage[T1]{fontenc}
\usepackage[latin9]{inputenc}
\usepackage[letterpaper]{geometry}
\usepackage{color}
\usepackage{float}
\usepackage{amsthm}
\usepackage{amssymb}
\usepackage{graphicx}

\makeatletter

\providecommand{\tabularnewline}{\\}
\floatstyle{ruled}
\newfloat{algorithm}{tbp}{loa}
\providecommand{\algorithmname}{Algorithm}
\floatname{algorithm}{\protect\algorithmname}

\numberwithin{equation}{section}
\theoremstyle{plain}
\newtheorem{thm}{\protect\theoremname}[section]
  \theoremstyle{definition}
  \newtheorem{example}[thm]{\protect\examplename}

\usepackage{ifpdf} 
\ifpdf 
 \IfFileExists{lmodern.sty}{\usepackage{lmodern}}{}
\fi 
\usepackage{color}

\usepackage{fancyhdr}
\fancypagestyle{plain}{\fancyfoot{}\fancyhead[C]{\text{\thepage}}}
\usepackage{graphicx}

\usepackage{hyperref}
\makeatother

\usepackage{babel}
  \providecommand{\examplename}{Example}
\providecommand{\theoremname}{Theorem}

\begin{document}

\title{Parameter Selection Algorithm For Continuous Variables}

\author{Peyman Tavallali$^{1,*}$, Marianne Razavi$^{1}$, Sean Brady$^{2}$}

\maketitle
$^{1}$Division of Engineering and Applied Sciences, California Institute
of Technology, 1200 East California Boulevard, MC 205-45, Pasadena,
CA 91125, USA

$^{2}$Principium Consulting, LLC, 556 S. Fair Oaks Ave., Ste. 101-264,
Pasadena, CA 91105

$^{*}$Corresponding Author, email: ptavalla@caltech.edu
\begin{abstract}
In this article, we propose a new algorithm for supervised learning
methods, by which one can both capture the non-linearity in data and
also find the best subset model. To produce an enhanced subset of
the original variables, an ideal selection method should have the
potential of adding a supplementary level of regression analysis that
would capture complex relationships in the data via mathematical transformation
of the predictors and exploration of synergistic effects of combined
variables. The method that we present here has the potential to produce
an optimal subset of variables, rendering the overall process of model
selection to be more efficient. The core objective of this paper is
to introduce a new estimation technique for the classical least square
regression framework. This new automatic variable transformation and
model selection method could offer an optimal and stable model that
minimizes the mean square error and variability, while combining all
possible subset selection methodology and including variable transformations
and interaction. Moreover, this novel method controls multicollinearity,
leading to an optimal set of explanatory variables.\textbf{ Keywords:}
Automatic Model Selection, Parameter and Variable Selection, Multivariate
Regression, Transformation and Interaction of Variables, Variance
Inflation Factor, Biostatistics 
\end{abstract}

\section{Introduction}

It happens often that the physical or mathematical model behind an
experiment or dataset is not known. Hence, model selection (sometimes
called subset selection) becomes an important feature in any data
analysis method. The methodology of selecting the best model has constantly
been examined by many authors. Identifying the best subset among many
variables is the most difficult part of model construction. The latter
is exacerbated as the number of possible subsets grows exponentially
if the number of variables (parameters) grows linearly. There is also
a chance that the original input parameters to a model do not convey
enough information, and some transformations of the original parameters
are needed to make the data more available for information extraction. 

In other words, in a supervised learning terminology, there is a long
and unpaved journey between the \emph{inputs} (also called \emph{predictors,
features} or \emph{independent variables}) and the \emph{outputs}
(also called \emph{responses} or \emph{dependent variables}). Thus,
the difficulty is not only embedded in picking the right machine learning
algorithm for the problem at hand, but also in picking proper transformations
of the inputs and their subsets. There are different methods capable
of addressing transformations and subset selection. However, to the
best of our knowledge, none of these methods solves both issues simultaneously.

In our discussions in this paper, we denote an input variable by an
$N\times1$ vector $\mathbf{x}$ as a collection of $N$ observations
of an input $x$. The assembly of $p$ such inputs and an intercept
is denoted by an $N\times\left(p+1\right)$ matrix $\mathbf{X}=\left(\mathbf{1},\mathbf{x}_{1},\mathbf{x}_{2},\ldots,\mathbf{x}_{p}\right)$.
The output is denoted by an $N\times1$ vector $\mathbf{Y}$. For
example, based on this description, a linear model is defined as 
\begin{equation}
\mathbf{Y}=\mathbf{X}\beta+\varepsilon,\label{eq: Linear Model}
\end{equation}
where $\varepsilon$ is the $N\times1$ noise vector, and $\beta=\left(\beta_{0},\beta_{1},\ldots,\beta_{p}\right)^{T}$
is a $\left(p+1\right)\times1$ vector of coefficients with the first
element being the \emph{intercept} (or \emph{bias}) of the model. 

In what follows next, we review a series of methods and algorithms
are used to find some subset(s) of the inputs that could possibly
relate the inputs to outputs in an efficient way.

\subsection{Subset Selection}

There are currently various methods for selecting predictors, such
as the traditional best subset selection, forward selection, backward
selection and stepwise selection methods \cite{friedman2001elements,montgomery2015introduction}.
In general, the best subset procedure finds for each $k\in\left\{ 1,2,\ldots,p\right\} $,
the subset of inputs of size $k$ that minimizes the Residual Sum
of Squares (RSS) \cite{furnival1971all,garside1965best,morgan1972calculation,schatzoff1968efficient}.
There are fast algorithms optimizing the search \cite{furnival2000regressions}.
However, searching through all possible subsets could become laborious
as $p$ increases. 

A number of automatic subset selection methods seek a subset of all
inputs, that is as close as possible to the best subset method \cite{montgomery2015introduction}.
These methods select a subset of predictors by an automated algorithm
that meets a predefined criterion, such as the level of significance
(set by the analyst). For example, the forward selection method \cite{montgomery2015introduction}
starts with no predictors in the model. It then adds predictors one
at a time until no available predictors can contribute significantly
to the response variable. Once a predictor is included in the model,
it remains there. On the other hand, the backward elimination technique
\cite{montgomery2015introduction} works in the opposite direction
and begins with all the existing predictors in the model, then discards
them one after another until all remaining predictors contribute significantly
to the response variable. Stepwise subset selection \cite{efroymson1960multiple}
is a mixture of the forward and backward selection methods. It modifies
the forward selection approach in that variables already in the model
do not always remain in the model. Indeed, after each step in which
a variable is added, all variables in the model are reevaluated via
their partial $F$ or $t$ statistics and any non-significant variable
is removed from the model. The stepwise regression requires two cutoff
values for significance: one for adding variables and one for discarding
variables. In general the probability threshold for adding variables
should be smaller than the probability threshold for eliminating variables
\cite{montgomery2015introduction}. 

Subset selection methods are usually based on targeting models with
the largest $R_{adj}^{2}$, or in other words smallest Root Mean Square
Error (RMSE). However, there are other methods in which the selection
model is based on Mallow's $C_{p}$ \cite{mallows1964choosing,mallows1995more,mallows1973some,mallows1967choosing}.
These criteria highlight different aspects of the regression model.
As a results, they can lead to models that are completely different
from each other and yet not optimal.

\subsection{Ridge Regression}

There are also other issues regarding the traditional subset selection
regression methods. They could lead to models that are unreliable
for prediction because of over-fitting issues. More specifically,
they could generate models that have variables displaying a high degree
of multicollinearity. Such methods can lead to $R^{2}$ values that
are biased and yield to confidence limits that are far too narrow
or wide. Moreover, the selection criterion primarily relies on the
correlation between the predictor(s) and the dependent variable. Thus,
these methods (e.g. Stepwise method \cite{olusegun2015identifying})
do not take into consideration the correlation within the predictors
themselves. The latter is a source of multicollinearity that is not
addressed automatically by these mentioned methods \cite{olusegun2015identifying}.

Indeed when collinearity among the predictors exists, the variance
of the coefficients is inflated, rendering the overall regression
equation unstable. To address this issue, a number of \emph{penalized
regression} or \emph{shrinkage} approaches are available. For example,
the Ridge method tries to eliminate the multicollinearity by imposing
a penalty on the size of the regression coefficients \cite{friedman2001elements}.
Indeed, a model is fitted with all the predictors, however, the estimated
coefficients are shrunken towards zero relative to the least squared
estimates. Therefore, biased estimators of regression coefficients
are obtained, reducing the variance and thus leading to a more stable
equation. 

Solving for $\beta$ in Equation (\ref{eq: Linear Model}) using the
Least Squares (LS) method would be equivalent to solving 
\begin{equation}
\hat{\beta}^{LS}=\underset{\beta}{\arg\min}\left(\mathbf{Y}-\mathbf{X}\beta\right)^{T}\left(\mathbf{Y}-\mathbf{X}\beta\right).\label{eq: Ordinary Least Squares}
\end{equation}
Ridge regression, on the other hand, places a constraint on the estimator
$\beta$ in order to minimize a penalized sum of squares \cite{hoerl1970ridgeA,hoerl1970ridgeB}
\begin{equation}
\hat{\beta}^{ridge}=\underset{\beta}{\arg\min}\left(\mathbf{Y}-\mathbf{X}\beta\right)^{T}\left(\mathbf{Y}-\mathbf{X}\beta\right)+\lambda\beta^{T}\beta.\label{eq: Ridge Least Squares}
\end{equation}
The complexity parameter $\lambda\geqq0$ controls the amount of shrinkage.
Large values of this parameter would result in a large shrinkage.
The value of the constant $\lambda$ is predefined by the analyst
and is usually selected in order to stabilize the ridge estimators,
producing an improved equation with a smaller RMSE compared to the
least-squares estimates. One weakness of the Ridge method is that
it does not select variables. Indeed, unlike the subset selection
method, it includes all of the predictors in the final model with
shrunken coefficients. The other weakness is that multicollinearity
is not addressed. In fact, the Ridge estimate of variables in (\ref{eq: Ridge Least Squares})
only shrinks the coefficients even for the inputs with multicollinearity.
However, the Ridge Method does not fix multicollinearity, it only
alleviates it.

\subsection{Lasso}

To obtain variable selection procedures, there are shrinkage methods
available such as Least Absolute Shrinkage and Selection Operator
(Lasso), where the penalty involves the sum of the absolute values
of the coefficients $\beta$ excluding the intercept \cite{chen1998atomic}.
Lasso is closely related to sparse optimization found in works by
Candes and Tao \cite{candes2006near}. Taking $\beta^{-}=\left(\beta_{1},\ldots,\beta_{p}\right)^{T}$,
the Lasso method can be presented as the following optimization problem
\begin{equation}
\hat{\beta}^{Lasso}=\underset{\beta}{\arg\min}\left(\mathbf{Y}-\mathbf{X}\beta\right)^{T}\left(\mathbf{Y}-\mathbf{X}\beta\right)+\lambda\left\Vert \beta^{-}\right\Vert _{1},\label{eq: Lasso}
\end{equation}
where $\left\Vert \beta^{-}\right\Vert _{1}=\sum_{1}^{p}\left|\beta_{j}\right|$
is the $L_{1}$ norm of $\beta^{-}$ and $\lambda>0$. The advantage
of Lasso is that much like the best subset selection method, it performs
variable selection. 

The parameter $\lambda$ is usually selected by cross validation.
For a small $\lambda$, the result is equal to the least squares estimates.
As the value of $\lambda$ augments, shrinkage happens in such a way
that only a sparse number of variables having an active role in the
final model would show up. Thus, Lasso is a combination of both shrinkage
and variables selection.

\subsection{LAR}

Least Angle Regression (LAR) is a new model of automatic subset selection
based on a modified version of forward procedure \cite{efron2004least}.
The LAR method follows an algorithmic procedure: First, the independent
variables are standardized in order to obtain a mean zero. At this
stage, the $\beta$ coefficients are all equal to zero. Then the predictor
that most correlates to the response variable is selected; its coefficient
is then shifted from zero towards its least squares value. Now, once
a second predictor becomes as correlated with the existing residual
as the first predictor, the procedure is paused. The second predictor
is then added to the model. This procedure then continues until all
desired predictors are included in the model, leading to a full least-squares
fit. 

The method of Least Angle Regression with Lasso modification is very
similar to the above procedure, however it includes an extra step:
if a coefficient approaches zero, LAR excludes its predictor from
the model and recalculates the joint least squares path \cite{friedman2001elements}.
LAR methods and its variations are better subset selector algorithms
compared to most of the subset selection methods.

\subsection{Dantzig}

Another selection approach is the Dantzig selector \cite{candes2007dantzig},
which can be formulated as 
\begin{equation}
\underset{\beta}{\min}\left\Vert \mathbf{X}^{T}\left(\mathbf{Y}-\mathbf{X}\beta\right)\right\Vert _{\infty}
\end{equation}
subject to $\left\Vert \beta\right\Vert _{1}\leq t$. Here, $\left\Vert .\right\Vert _{\infty}$
is the $L_{\infty}$ norm, that is the maximum of its argument. The
objective of this method is to minimize the maximum inner product
of the existing residual with all the independent variables. This
approach has the capacity of recovering an underlying sparse coefficient
vector.

\subsection{PCR}

Lastly, Principal Component Regression (PCR) is a method that involves
an orthogonal transformation to address multicollinearity \cite{friedman2001elements,stone1990continuum,zou2006sparse}.
This approach is closely related to the Singular Value Decomposition
(SVD) method \cite{trefethen1997numerical}. PCR applies dimensionality
reduction and decreases multicollinearity by using a subset of the
principal components in the model \cite{friedman2001elements}. PCR
is one of very few methods that tries to eliminate multicollinearity
with linear transformations and, at the same time, perform a regression. 

The various approaches described earlier aim to select the best set
of relevant variables from an original set. With the exception of
the PCR method, in which there is linear transformations, variables
transformations are not incorporated among predictors in any of the
methods mentioned above. These traditional methods do not offer the
option of automatic variable transformation to address polynomial
curvilinear relationships. The analyst usually needs to manually apply
polynomial, logarithmic, square-root and interaction-between-variables
transformations in order to address non-linearity of the data.

\subsection{Non-Linear Transformation}

There are a number of non-linear transformation procedures currently
available such as Box-Cox or Box-Tidwell \cite{box1964analysis,box1962transformation}.
These methods are relatively efficient in finding the dependent and
independent variables transformations. In Box-Tidwell method \cite{box1962transformation},
independent variables are transformed using a recursive Newton algorithm.
As a result, it becomes susceptible to round-off errors which would
in turn result in unstable and improper transformations \cite{montgomery2015introduction}.
Despite the relative success of these methods, there is no automatic
variable selection embodiment with them.

Artificial Neural Networks (ANN) are the current state of the art
method in transformations and capturing non-linearity \cite{friedman2001elements,mackay2003information}.
ANN is a machine learning method that finds some non- linear transformations
of the inputs using layers of nodes. Despite the efficient performance
in capturing the non-linearity of the data, the model itself is not
comprehensible particularly if there is a physical component to the
data that one needs to understand. In other words, ANN is a perfect
black box model, but not a good interpretation medium for understanding
physical and mathematical mechanism(s) behind the observed data.

\subsection{Subset Selection and Transformation}

As mentioned earlier, only the PCR method performs linear transformations
automatically, and also picks variables. However, PCR is not enough
when non-linearity is present. On the other hand, ANN has the best
capability in capturing non-linearities, but acts like a black box
and does not lend insight into the physical and mathematical mechanism(s)
behind the observed data.

To produce an enhanced subset of the original variables, an ideal
selection method should have the potential of adding a supplementary
level of regression analysis that would capture complex relationships
in the data via mathematical transformation of the predictors and
exploration of synergistic effects of combined variables. The method
that we present here has the potential to produce an optimal subset
of variables, resulting in a more efficient overall process of model
selection.

The core objective of this paper is to introduce a new estimation
technique for the classical least square regression framework. This
new automatic variable transformation and model selection method could
offer an optimal and stable model that minimizes the mean square error
and variability, while combining all possible subset selection methodologies
and including variable transformations and interaction. Moreover,
this novel method controls multicollinearity, leading to an optimal
set of explanatory variables. The resultant model is also easy to
interpret. In other words, we will depict a method that addresses
variable selection and transformation at the same time; and also helps
the analyst make interpretations about the physical and mathematical
mechanism(s) behind the observed data.

\section{Methodology}

As mentioned in the previous section, the analyst usually needs to
manually apply polynomial, logarithmic, square-root and interaction
between variables in order to address the non-linearity of the data.
At the same time, the analyst should also look for the best model
from a subset of the variables. However, this manual burden can be
made automatic. We will first address the issue of transformations
and subset selection separately and then put them together. 

There are four important categories of transformations to capture
the non-linearity in a data set \cite{friedman2001elements}. These
transformations are
\begin{enumerate}
\item Logarithm transformation of a positive variable; i.e. $\log x_{j}$
\item Square-root transformation of a positive variable; i.e. $\sqrt{x_{j}}$
\item Integer powers up to a certain amount $\alpha\mathbb{\in N}$; i.e.
$\left\{ \frac{1}{x_{j}^{\alpha}},\frac{1}{x_{j}^{\alpha-1}},\ldots,x_{j}^{\alpha-1},x_{j}^{\alpha}\right\} $
\item Interaction between terms created in 1-3 to a certain mixture number
$M$; e.g., for $M=2$, possible candidates for would be $\frac{1}{x_{i}}$,
$\frac{x_{i}^{2}}{x_{j}}$, $x_{i}$, $x_{i}^{2}x_{j}^{2}$, $x_{i}^{2}\left(\log x_{j}\right)^{2}$
and $\frac{\sqrt{x_{i}}}{x_{j}}$.
\end{enumerate}
After the construction of these transformations, one can start to
look for the best model, for $\mathbf{Y}$, among the set of all mentioned
transformations 1-4. In other words, denoting the set of variables
created by transformations 1-4 as $\mathbf{Z}$, we are looking for
the best sparse model 
\begin{equation}
\mathbf{Y}=\mathbf{Z}\beta_{z}+\varepsilon,\label{eq: Transformation Raw}
\end{equation}
where some elements of $\beta_{z}$ are zero. In fact, some elements
of $\beta_{z}$ are zero because there is a chance that some columns
of $\mathbf{Z}$ are linearly dependent or that they do not contribute
to any correlation to $\mathbf{Y}$. 

Up to this stage, we see that using transformations, we can expand
the basis of projection. This would help us to find better correlations
between transformed variables and the output $\mathbf{Y}$. In other
words, using these transformations, we essentially convert the problem
into a dictionary search \cite{chen1998atomic}. As we mentioned before
some of the transformations might not be of any importance in (\ref{eq: Transformation Raw}),
or could be redundant due to multicollinearity. We can address these
two issues, by a modified dictionary search algorithm, as follows:
\begin{itemize}
\item Any column of $\mathbf{Z}$ that has a non-significant correlation
(less than $\delta$) with $\mathbf{Y}$ can be discarded. In other
words, the corresponding elements of $\beta_{z}$, having a small
importance, will be set to zero. 
\item Any two columns of $\mathbf{Z}$ that have a high correlation with
each other (greater than $\varsigma$) are redundant columns. Between
these two, one should pick the one that has a higher correlation with
$\mathbf{Y}$ and discard the other. In other words, the element in
$\beta_{z}$ that corresponds to a redundancy with a weaker correlation
to $\mathbf{Y}$ can be set to zero.
\end{itemize}
As a result of this methodology, we can now solve model (\ref{eq: Transformation Raw})
for only a reduced matrix. Hence, although we construct a matrix dictionary
of transformations $\mathbf{Z}$ that is larger than the original
input matrix $\mathbf{X}$, we can reduce the size of $\mathbf{Z}$
as some elements of $\beta_{z}$ are now set to zero. We denote this
reduced matrix as $\mathbf{Z}^{r}$ and its corresponding vector of
coefficients as $\beta_{z}^{r}$. This reduction is similar to Basis
Pursuit (BP) method presented in \cite{chen1998atomic}. 

After reducing the matrix of transformations, the final task is to
find the best subset of the columns in $\mathbf{Z}^{r}$ to model
the data in $\mathbf{Y}$. The latter can be done by any method of
subset selection; e.g. best subset selection. For simplicity in referencing
to our proposed method, we denote the whole procedure as Parameter
Selection Algorithm (PARSEAL). In the next section, we formalize the
methodology that we explained here.

\section{PARSEAL}

The goal of the PARSEAL is to find the best model explaining $\mathbf{Y}$
from some important transformations on the original observed variables
$\mathbf{X}$. The latter is nothing but finding the best subspace
using some specific transformations from the original inputs. 

The PARSEAL is presented in Algorithm \ref{Alg: PARSEAL}. This algorithm
is equipped with the best subset selection method to find the best
model by maximizing the $R_{adj}^{2}$ among all possible subset of
variables in $\mathbf{Z}^{r}$. Step 1 of this algorithm is input
specification. Step 2 is where the dictionary of transformations and
interactions is made. Steps 3 and 4 correspond to the elimination
of columns of the dictionary of transformations and interactions which
involve either a non-significant correlation to the output or multicollinearity
between the dictionary elements. Step 5 is where the best model is
finally found. Steps 2, 3 and 5 in this algorithm can be made parallel
to decrease the computational time of the method. To our best knowledge,
Algorithm \ref{Alg: PARSEAL} is the first method that performs both
variable transformation and model selection while adding interaction
terms and also preventing multi-collinearity, in one package. 

The importance limit $\delta$ and independence limit $\varsigma$
are important factors in controlling the speed of convergence of the
PARSEAL. In Algorithm \ref{Alg: PARSEAL}, the smaller the value of
$\delta$ (similarly, the larger the value of $\varsigma$), the bigger
the space of search in step 5. As a result, the speed of convergence
would depend greatly on these two parameters. 

In step 5, checking for Variation Inflation Factors (VIFs) \cite{marquaridt1970generalized}
is a necessary condition to make sure that no multicollinearity is
introduced into the final model. VIFs are the elements of the main
diagonal of the inverse of the multiplication of the input matrix
transposed with the input matrix. For example if $\mathbf{X}$ is
the input, then $\mathbf{C}=\left(\mathbf{X}^{T}\mathbf{X}\right)^{-1}$
and $VIF_{j}=C_{jj}$. Checking the VIF is an essential multicollinearity
diagnosis tool \cite{montgomery2015introduction}.

\begin{algorithm}[H]
\begin{enumerate}
\item Inputs to the algorithm: 

\begin{enumerate}
\item Original input matrix $\mathbf{X}$
\item Output vector $\mathbf{Y}$
\item Maximum polynomial power $\alpha$
\item Maximum term interaction number $M$
\item Importance limit $\delta$
\item Independence limit $\varsigma$
\end{enumerate}
\item Construct matrix $\mathbf{Z}$ from $\mathbf{X}$ by taking logarithm,
square-root, polynomial and interactions.
\item Eliminate columns of $\mathbf{Z}$ having an absolute-value correlation
less than the importance limit $\delta$: $\mathbf{Z}^{1}$
\item For any two columns of $\mathbf{Z}^{1}$ having an absolute-value
correlation greater than the independence limit $\varsigma$, pick
the column that has a higher absolute-value correlation with $\mathbf{Y}$:
$\mathbf{Z}^{r}$ 
\item Find the best subset of variables in $\mathbf{Z}^{r}$ against $\mathbf{Y}$
by maximizing $R_{adj}^{2}$ while preventing the VIF of any variable
exceeding 10.
\end{enumerate}
\caption{PARSEAL}

\label{Alg: PARSEAL}
\end{algorithm}

\subsection{Values for $\varsigma$ }

In fact, the independence limit $\varsigma$ can be characterized
with the VIF concept. As mentioned, $VIF_{j}=C_{jj}$, however, this
formula can also be written as 
\[
VIF_{j}=C_{jj}=\frac{1}{1-R_{j}^{2}}.
\]
 Here, $R_{j}^{2}$ is the multiple $R^{2}$ for the regression of
$X_{j}$ against other inputs. Hence, if we want two inputs to have
a small correlation with each other, we must have a possible VIF between
them to be less than 10. This would impose an $R^{2}=0.9$ between
those variables. Hence, a correlation of $\sim0.95$ would say if
two inputs are highly correlated or not. On the other hand, we know
that if we set the independence limit $\varsigma=0.95$, we would
construct a huge dictionary of inputs when transformations are available.
Hence, based on our numerical experiments, a value of $\varsigma=0.80$
is more practical.

\section{Synthetic Examples}

\begin{figure}[H]
\includegraphics[angle=-90,scale=0.4]{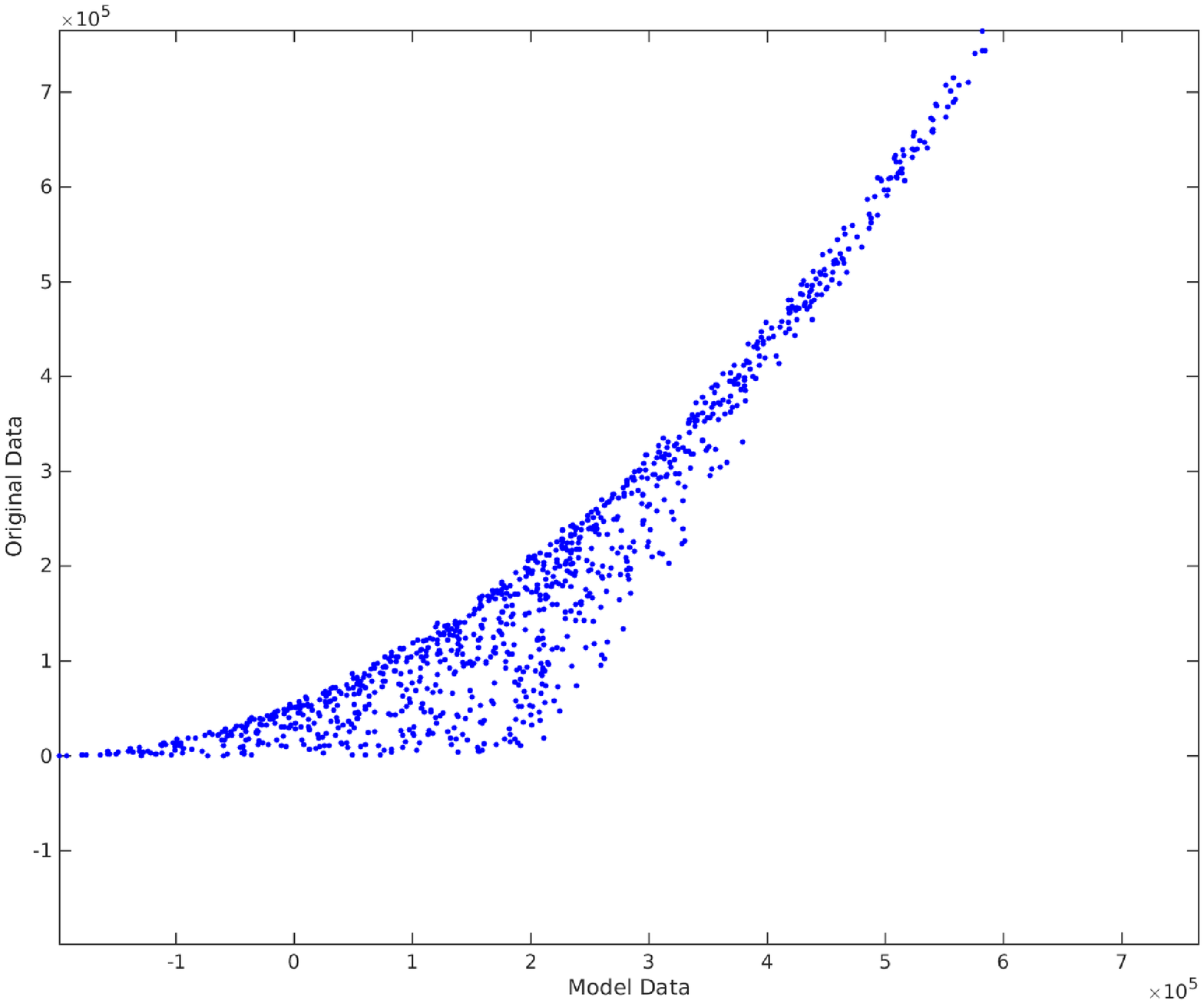}

\caption{Traditional best subset selection method applied on $y=120+80x_{1}x_{3}$.
The horizontal axis shows the model found by the best subset selection
method. The vertical axis shows the output $y$.}

\label{Fig: Ex1 Original Data vs Model Data}
\end{figure}

\begin{figure}[H]
\includegraphics[angle=-90,scale=0.4]{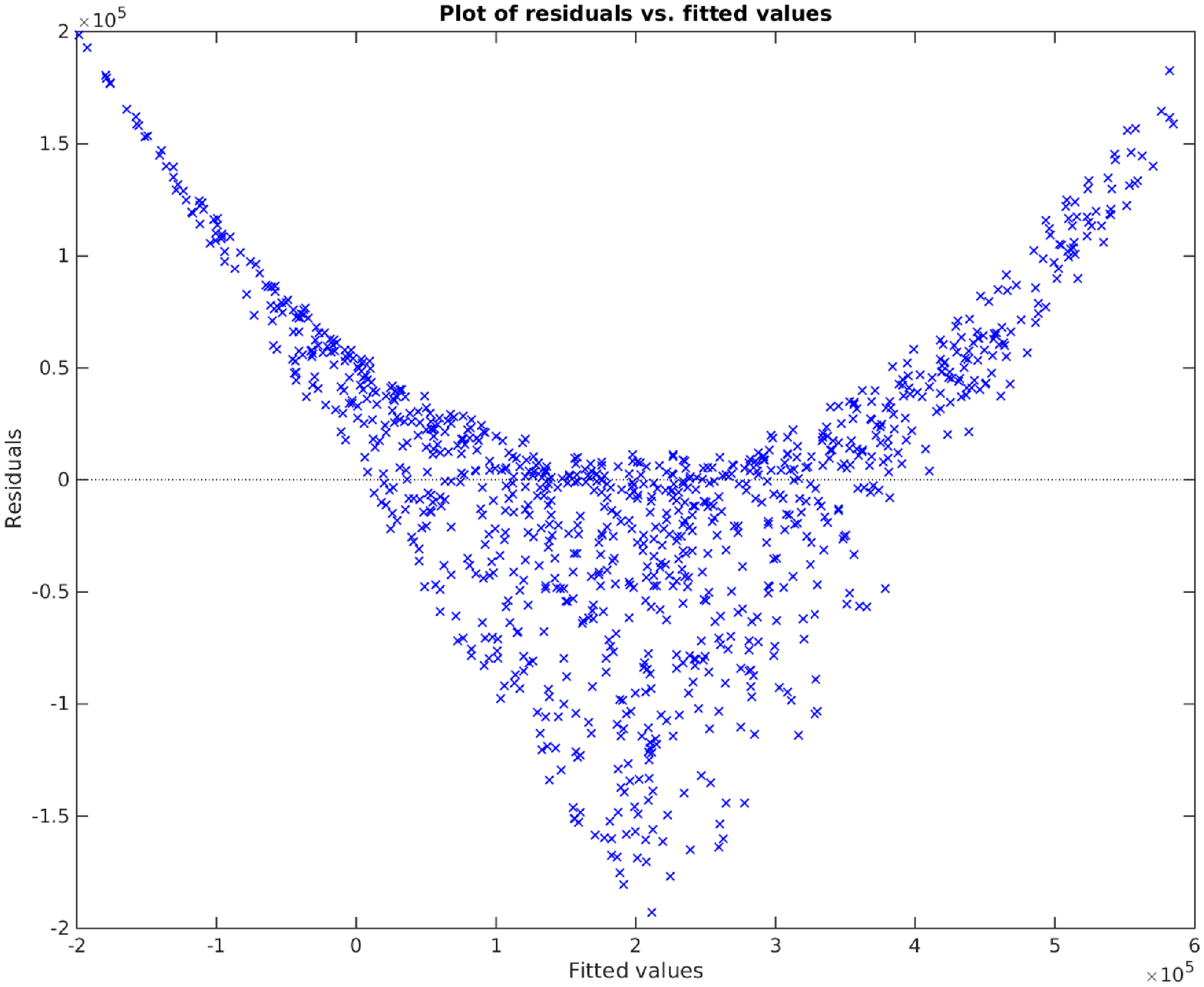}

\caption{Residual plot of the best subset selection method applied on $y=120+80x_{1}x_{3}$.}

\label{Fig: Ex1 Residuals}
\end{figure}

\begin{figure}[H]
\includegraphics[angle=-90,scale=0.4]{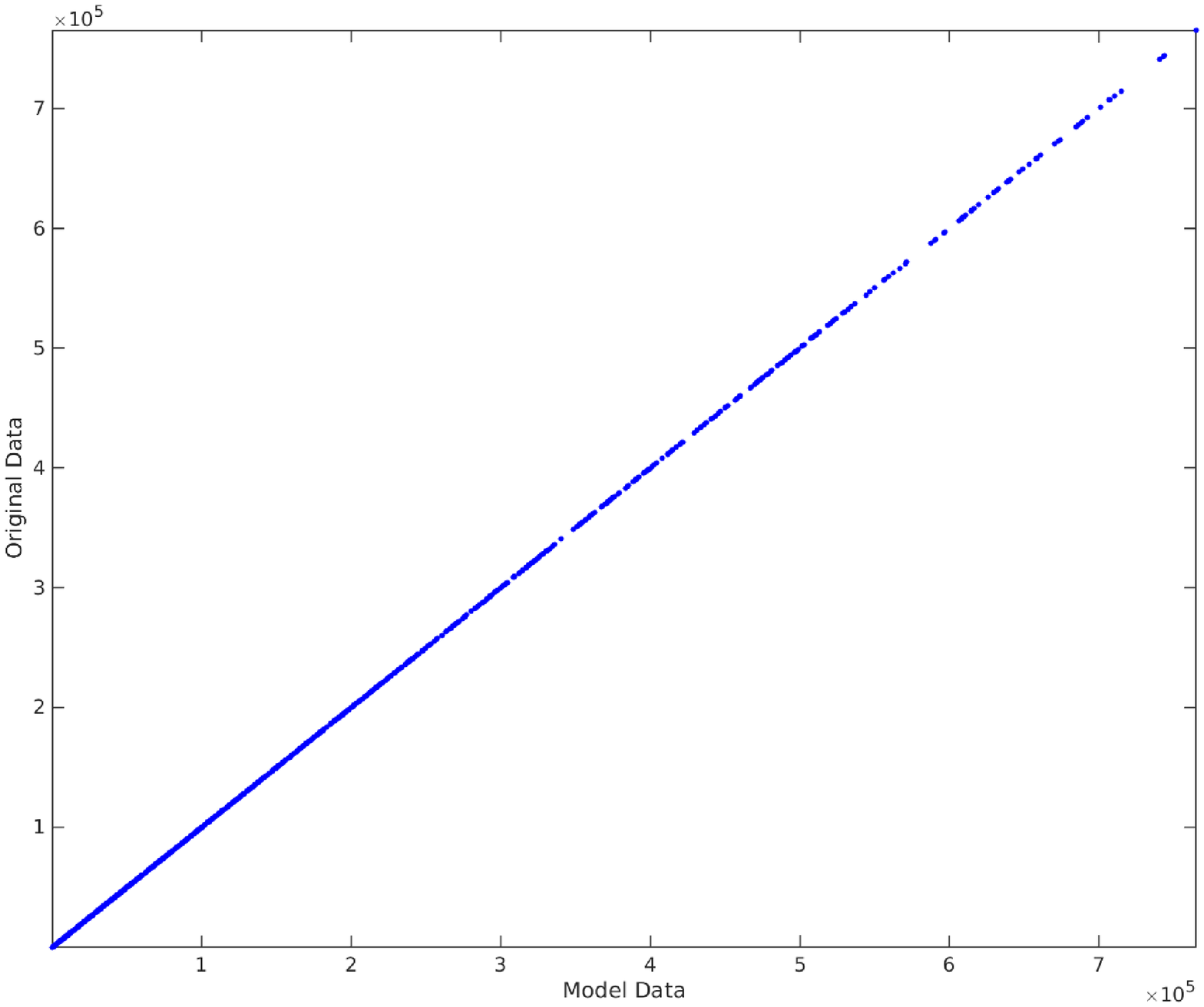}

\caption{Algorithm \ref{Alg: PARSEAL} applied on $y=120+80x_{1}x_{3}$. The
horizontal axis shows the model found by our proposed method. The
vertical axis shows the output $y$.}

\label{Fig: Ex1 Original Data vs Model Data Parseal}
\end{figure}

\begin{figure}[H]
\includegraphics[angle=-90,scale=0.4]{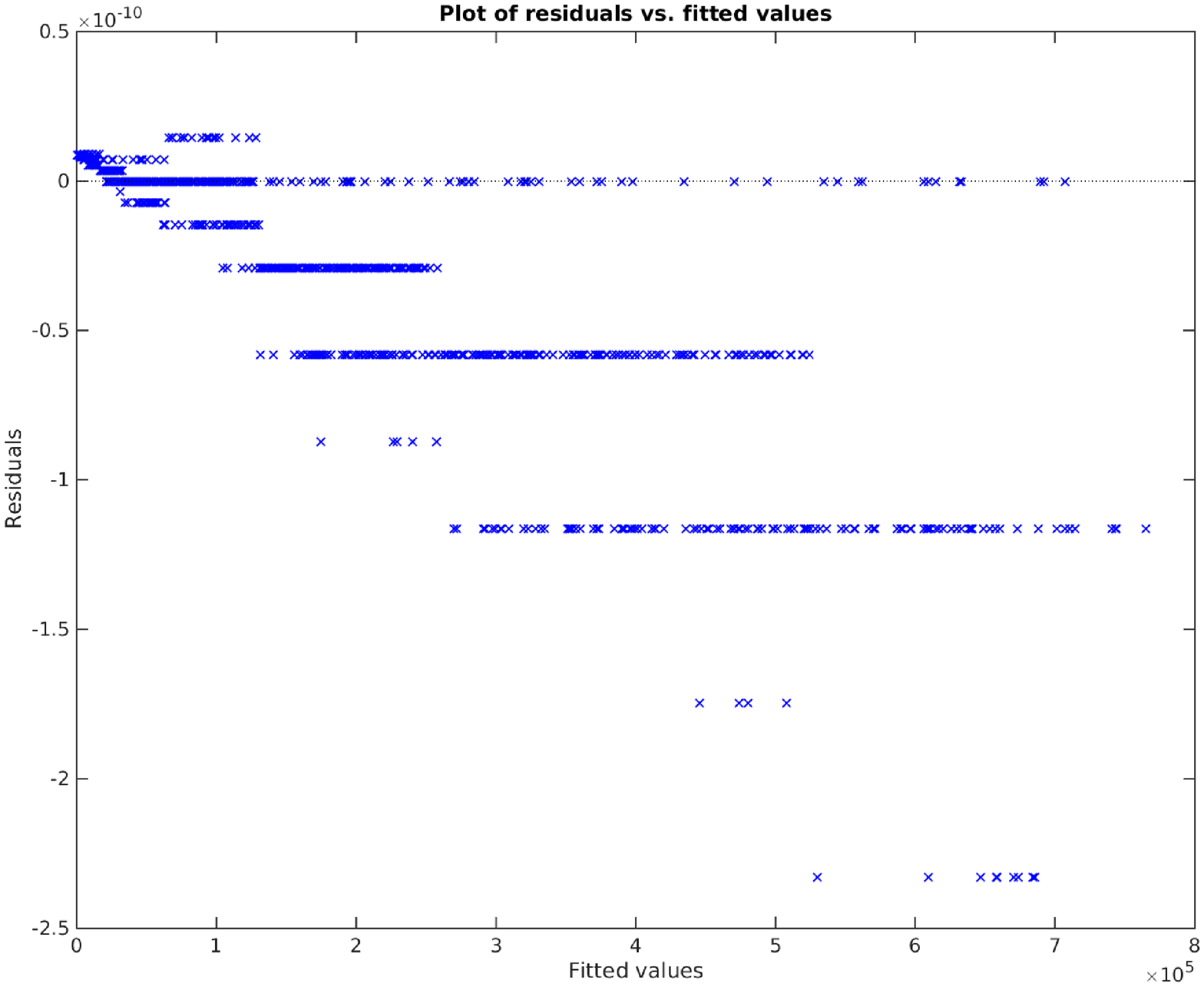}

\caption{Residual plot of our proposed method applied on $y=120+80x_{1}x_{3}$.
Note that the vertical axis is of the order $10{}^{-10}$. The error
perceived here is due to floating point and rounding error.}

\label{Fig: Ex1 Residuals for Parseal}
\end{figure}

In this section, we provide a few synthetic examples using PARSEAL.
In the following examples, we try to show that the algorithm that
we have proposed is capable of finding the non-linear transformations
in a model.
\begin{example}
Taking $x_{1}$, $x_{2}$, $x_{3}$ to be a uniformly distributed
random variable in $\left[0,100\right],$ we sampled $1000$ data
points and then created the non-linear functional $y=120+80x_{1}x_{3}$.
We take the original input matrix $\mathbf{X}$ to be composed of
all $x_{1}$, $x_{2}$, and $x_{3}$. Using the traditional best subset
selection \cite{furnival2000regressions}, accompanied with a control
over VIFs not to get above $10$, we get the results shown in Figure
\ref{Fig: Ex1 Original Data vs Model Data}. From this figure, it
is clear that the best subset selection model is not capable of capturing
the correct non-linearity in the model. The heteroscedasticity of
the residual plot can be seen from Figure \ref{Fig: Ex1 Residuals}.
The found subset of parameters is $\left\{ x_{1},x_{2},x_{3}\right\} $. 

On the other hand, if Algorithm \ref{Alg: PARSEAL} is used, the non-linearity
is captured completely by our method (See Figures \ref{Fig: Ex1 Original Data vs Model Data Parseal}
and \ref{Fig: Ex1 Residuals for Parseal}). The subset of parameters
found by our method is the model non-linear parameter $\left\{ x_{1}x_{3}\right\} $.
Here, the importance limit $\delta$ was half of the maximum absolute
value of the correlation of all columns in $Z$ compared to the output.
The independence limit was $\varsigma=0.5$.
\end{example}

\begin{figure}[H]
\includegraphics[angle=-90,scale=0.4]{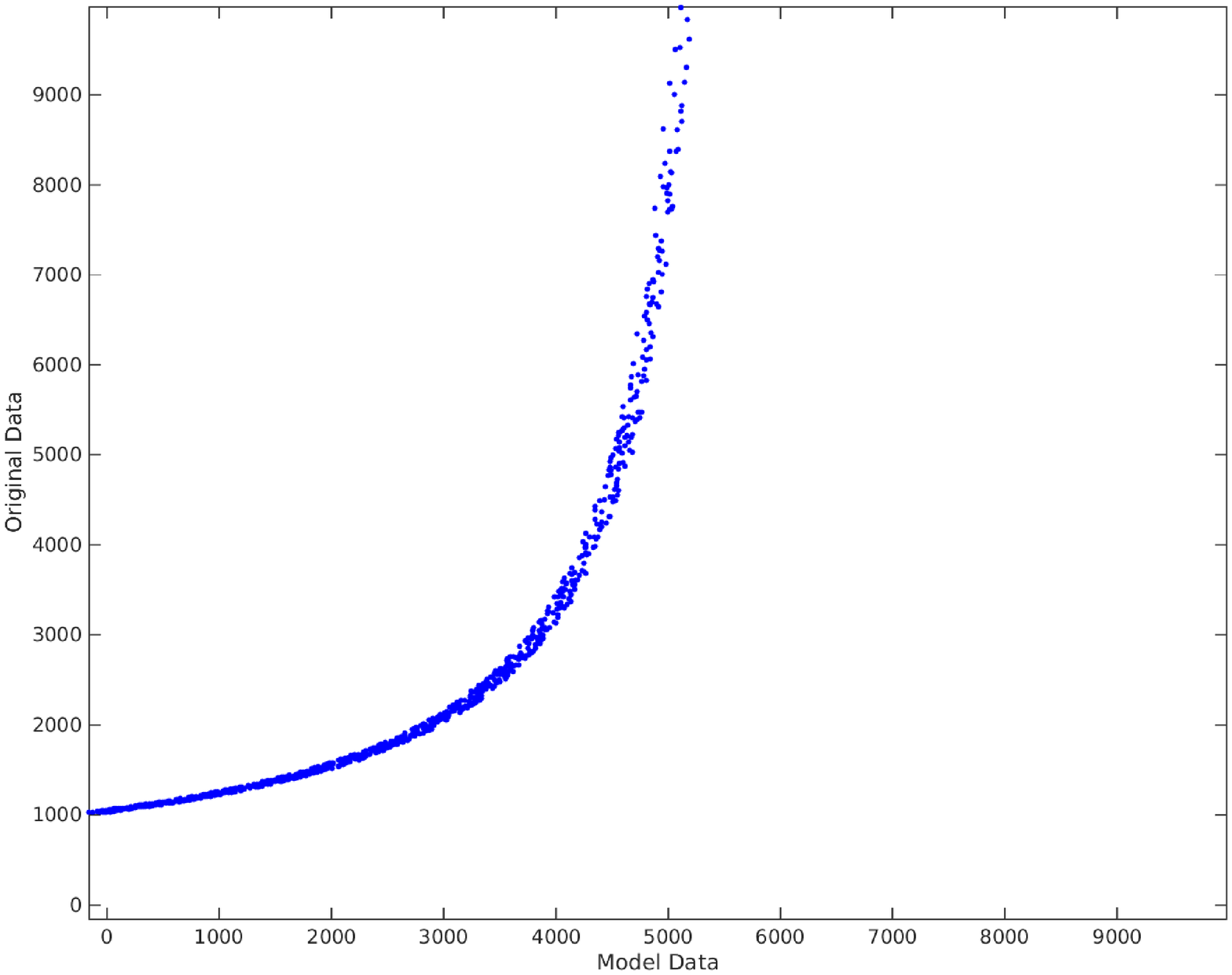}

\caption{Traditional best subset selection method applied on $y=120+\frac{1000}{x_{2}}$.
The horizontal axis shows the model found by the best subset selection
method. The vertical axis shows the output $y$.}

\label{Fig: Ex2 Original Data vs Model Data}
\end{figure}

\begin{figure}[H]
\includegraphics[angle=-90,scale=0.4]{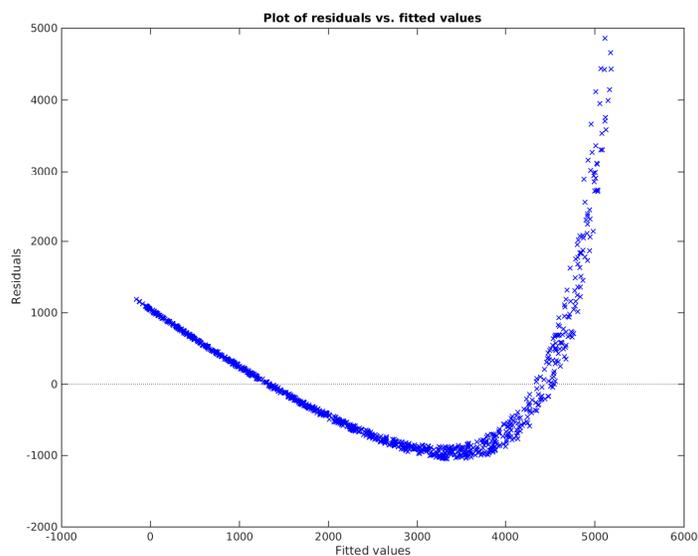}

\caption{Residual plot of the best subset selection method applied on $y=120+\frac{1000}{x_{2}}$.}

\label{Fig: Ex2 Residuals}
\end{figure}

\begin{figure}[H]
\includegraphics[angle=-90,scale=0.4]{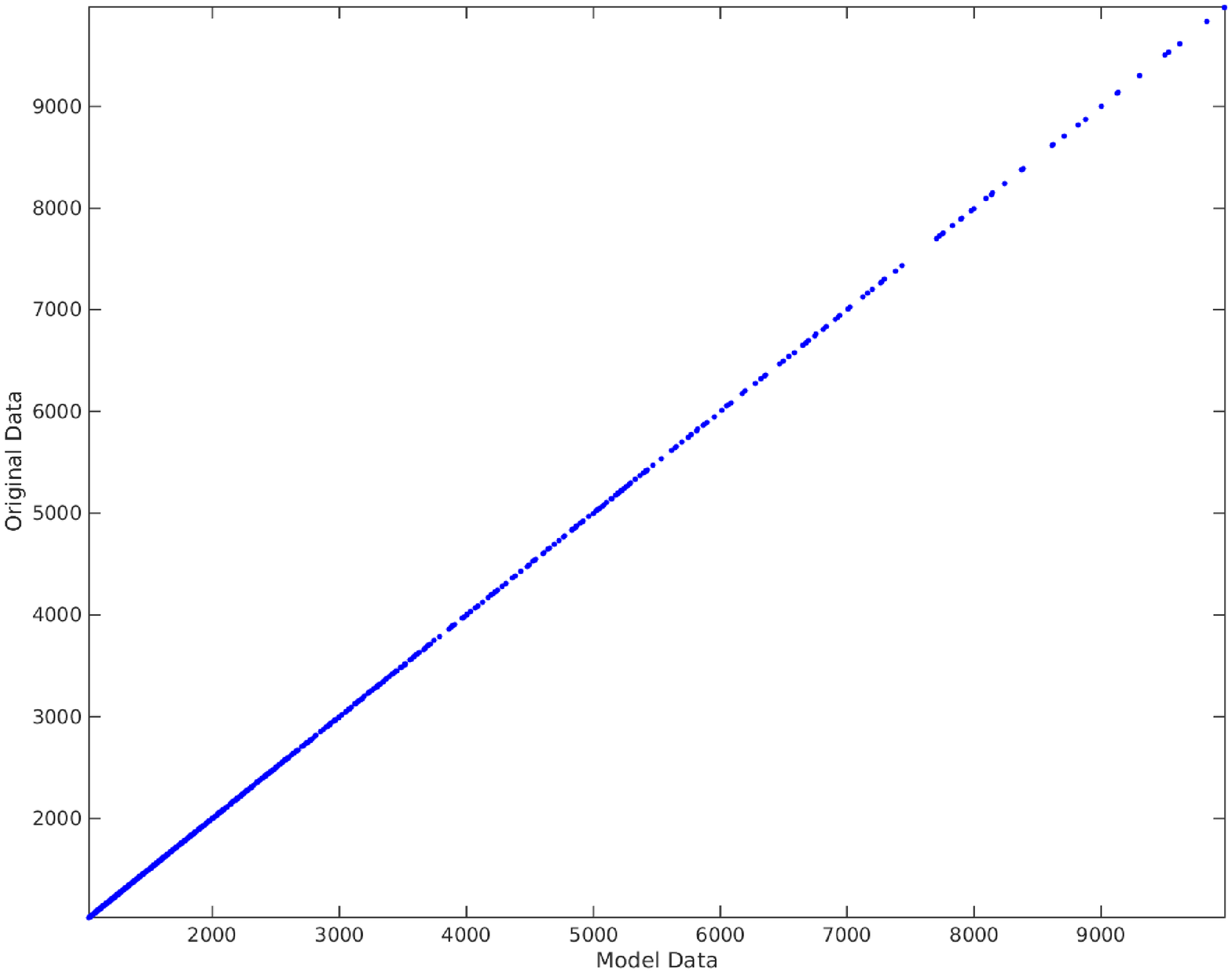}

\caption{Algorithm \ref{Alg: PARSEAL} applied on $y=120+\frac{1000}{x_{2}}$.
The horizontal axis shows the model found by our proposed method.
The vertical axis shows the output $y$.}

\label{Fig: Ex2 Original Data vs Model Data Parseal}
\end{figure}

\begin{figure}[H]
\includegraphics[angle=-90,scale=0.4]{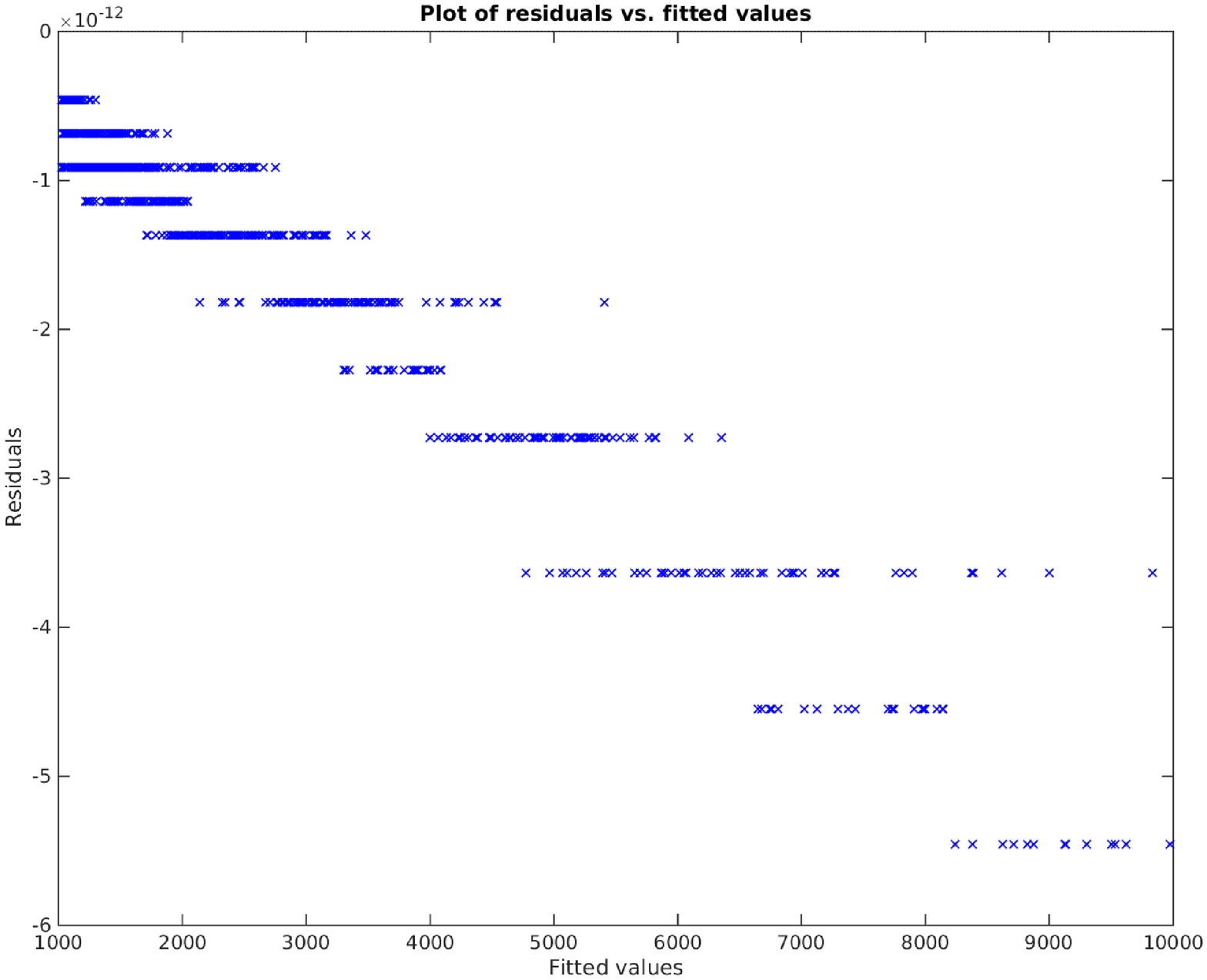}

\caption{Residual plot of our proposed method applied on $y=120+\frac{1000}{x_{2}}$.
Note that the vertical axis is of the order $10{}^{-12}$. The error
perceived here is due to floating point and rounding error.}

\label{Fig: Ex2 Residuals for Parseal}
\end{figure}

\begin{example}
If $\chi$ is a uniform random random variable in $\left[0,1\right]$,
then we set 
\[
\begin{array}{c}
x_{1}=100\chi,\\
x_{2}=\chi+0.1,\\
x_{3}=100\chi.
\end{array}
\]
We sampled $1000$ data points of $x_{1}$, $x_{2}$, and $x_{3}$
and then created the non-linear functional $y=120+\frac{1000}{x_{2}}$.
We take the original input matrix $\mathbf{X}$ to be composed of
all $x_{1}$, $x_{2}$, and $x_{3}$. Using the traditional best subset
selection \cite{furnival2000regressions}, accompanied with a control
over VIFs not to get above $10$, we get the results shown in Figure
\ref{Fig: Ex2 Original Data vs Model Data}. Again, from this figure,
it is clear that the best subset selection model is not capable of
capturing the correct non-linearity in the model. The heteroscedasticity
of the residual plot can be seen from Figure \ref{Fig: Ex2 Residuals}.
The found subset of parameters is $\left\{ x_{1},x_{2}\right\} $. 

On the other hand, if Algorithm \ref{Alg: PARSEAL} is used, the non-linearity
is captured completely (See Figures \ref{Fig: Ex2 Original Data vs Model Data Parseal}
and \ref{Fig: Ex2 Residuals for Parseal}). The subset of parameters
found by our proposed method is the model non-linear parameter $\left\{ \frac{1}{x_{2}}\right\} $.
Here, the importance limit $\delta$ was half of the maximum absolute
value of the correlation of all columns in $Z$ compared to the output.
The independence limit was $\varsigma=0.8$.
\end{example}

\section{Real Data Example}

The synthetic examples in the previous section showed the capability
of our method in capturing the true non-linearity. In this section,
we show a real data case study.

Cardiovascular Diseases (CVDs) are the major cause of deaths in the
United States, killing more than 350,000 people every year \cite{american2015heart}.
One of the major contributors to CVDs is arterial stiffness \cite{mitchell2010arterial,mitchell2004changes}.
Arterial stiffness can be approximated by Carotid-femoral Pulse Wave
Velocity (PWV) \cite{safar2000therapeutic}. In fact, PWV is one of
the most important quantitative index for arterial stiffness \cite{mitchell2004changes}.
PWV measures the speed of the arterial pressure waves traveling along
the blood vessels and higher PWV usually highlights stiffer arteries.
Increased aortic stiffness is related to many clinically adverse cardiovascular
outcomes \cite{mitchell2010arterial}. PWV constitutes an independent
and valuable marker for cardiovascular diseases (CVDs) and its use
is crucial as a routine tool for clinical patient assessment.

In this section, our aim is not to present the most accurate PWV model.
However, our goal is to show that if our technique of model construction
is used (see Algorithm \ref{Alg: PARSEAL}), we are able to find a
more physically and statistically valid model. 

The data we present is collected from $5444$ Framingham Heart Study
(FHS) participants \cite{FHS}. Each participant had undergone an
arterial tonometry data collection. The participants were part of
FHS Cohorts Gen 3 Exam 1 \cite{splansky2007third}, Offspring Exam
7 \cite{kannel1979investigation}, and Original Exam 26 \cite{dawber1951epidemiological}.
Here, we try to find models for PWV based on the following inputs:
Age ($A$), Pulse Duration ($D$), Weight ($W$), Height ($H$), and
Body Mass Index ($BMI$). One model is based on the traditional best
subset selection method monitored for $VIF<10$, and the other based
on the PARSEAL method (Algorithm \ref{Alg: PARSEAL}). The participant
characteristics are shown in Table \ref{Tbl: Participant Char}.

\begin{table}[H]
\begin{tabular}{|c|c|c|}
\hline 
 & Range & Median\tabularnewline
\hline 
Duration & 0.58 to 1.77 & 0.98\tabularnewline
\hline 
Age & 19 to 99 & 46\tabularnewline
\hline 
Weight & 83 to 339 & 165\tabularnewline
\hline 
Height & 54.00 to 78.75 & 66.50\tabularnewline
\hline 
BMI & 15.47 to 51.47 & 25.89\tabularnewline
\hline 
PWV & 3.5 to 29.6 & 7.4\tabularnewline
\hline 
\end{tabular}

\caption{Participant Characteristics}

\label{Tbl: Participant Char}
\end{table}

\subsection{Best Subset Selection Model Results:}

\begin{figure}[H]
\includegraphics[angle=-90,scale=0.4]{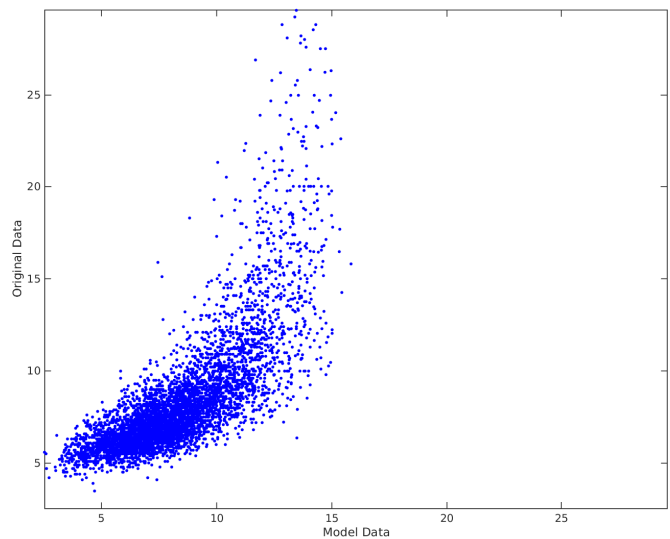}

\caption{Traditional best subset selection method applied on PWV data. The
horizontal axis shows the model found by the best subset selection
method. The vertical axis shows the recorded PWV data. The $R_{adj}^{2}$
of the model is $0.56737$.}

\label{Fig: Best Subset PWV Model}
\end{figure}

Figure \ref{Fig: Best Subset PWV Model} shows the traditional best
subset selection method applied on PWV data. As seen in the plot,
the best subset selection model cannot capture the non-linearity in
the data set and completely misses the PWV values above $15$. The
heteroscedasticity of the residual can be seen from the Bland-Altman
plot in Figure \ref{Fig: Best Subset PWV Model BA} and residual plot
in Figure \ref{Fig: Best Subset PWV Model Hete}. The $R_{adj}^{2}$
of this model is $0.56737$. The found subset of parameters is $\left\{ D,A,BMI,H\right\} $.
The p-value of these parameters are $3\times10^{-45}$, $0$, $2\times10^{-14}$,
and $1\times10^{-11}$, receptively.

\subsection{PARSEAL Results:}

Figure \ref{Fig: Parseal PWV Model} shows the PARSEAL method applied
on PWV data. Here, the importance limit $\delta$ was half of the
maximum absolute value of the correlation of all columns in $Z$ compared
to the output. The independence limit was $\varsigma=0.8$. As seen
on the plot, PARSEAL can fairly capture the non-linearity in the data
set. The residuals can be seen in the Bland-Altman plot in Figure
\ref{Fig: Parseal PWV Model BA} and residual plot in Figure \ref{Fig: Parseal PWV Model Hete}.
The $R_{adj}^{2}$ of the model is $0.63052$. The found subset of
parameters is $\left\{ DA,\frac{A^{2}}{\sqrt{D}},\frac{\left(\log\left(A\right)\right)^{2}}{\left(\log\left(W\right)\right)^{2}},\frac{\left(\log\left(BMI\right)\right)^{2}}{A}\right\} $.
The p-value of these parameters are $6\times10^{-4}$, $0$, $2\times10^{-21}$,
and $8\times10^{-46}$, receptively.

\subsection{Comparison and Results Discussion:}

\begin{figure}[H]
\includegraphics[angle=-90,scale=0.4]{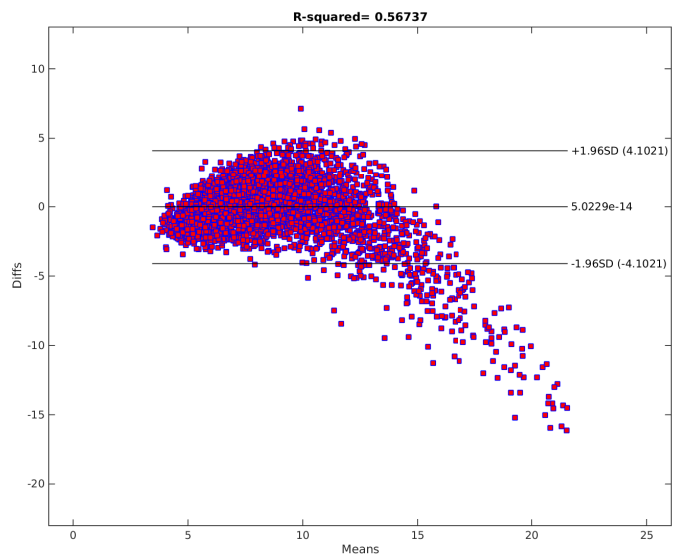}

\caption{Bland-Altman of the traditional best subset selection model. The horizontal
axis shows the means of the fitted and original PWV values. The Vertical
axis shows the differences between the fitted and original PWV values. }

\label{Fig: Best Subset PWV Model BA}
\end{figure}

\begin{figure}[H]
\includegraphics[angle=-90,scale=0.4]{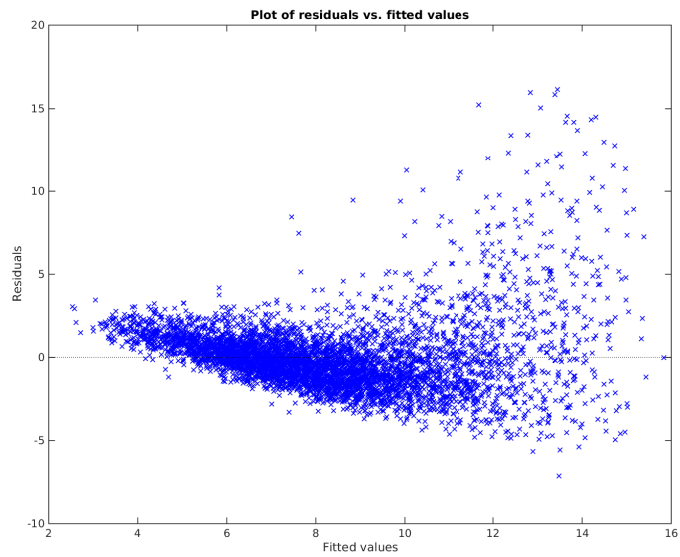}

\caption{Residual plot of best subset selection method.}

\label{Fig: Best Subset PWV Model Hete}
\end{figure}

\begin{figure}[H]
\includegraphics[angle=-90,scale=0.4]{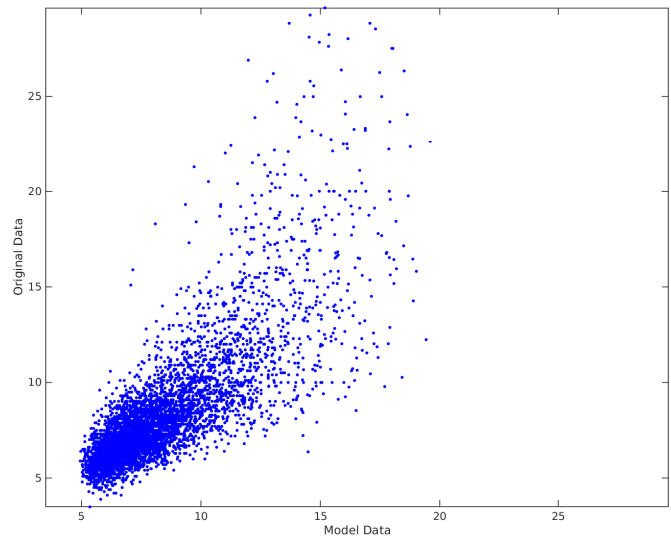}

\caption{PARSEAL method applied on PWV data. The horizontal axis shows the
model found by the best subset selection method. The vertical axis
shows the recorded PWV data. The $R_{adj}^{2}$ of the model is $0.63052$.}

\label{Fig: Parseal PWV Model}
\end{figure}

\begin{figure}[H]
\includegraphics[angle=-90,scale=0.4]{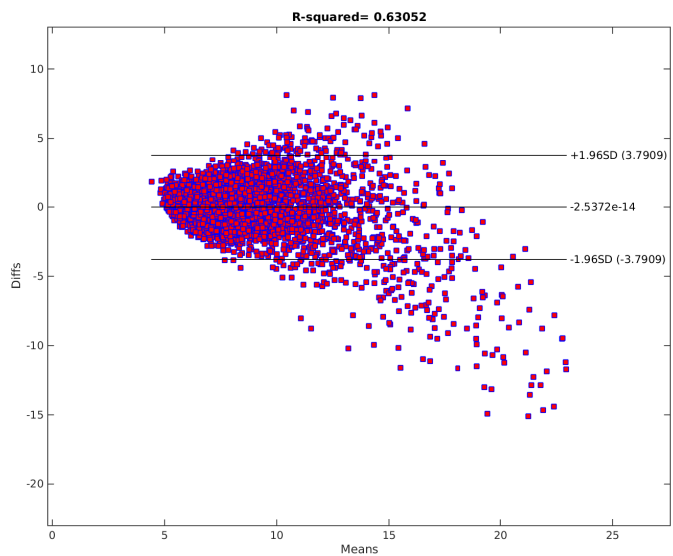}

\caption{Bland-Altman of the PARSEAL model. The horizontal axis shows the means
of the fitted and original PWV values. The Vertical axis shows the
differences between the fitted and original PWV values.}

\label{Fig: Parseal PWV Model BA}
\end{figure}

\begin{figure}[H]
\includegraphics[angle=-90,scale=0.4]{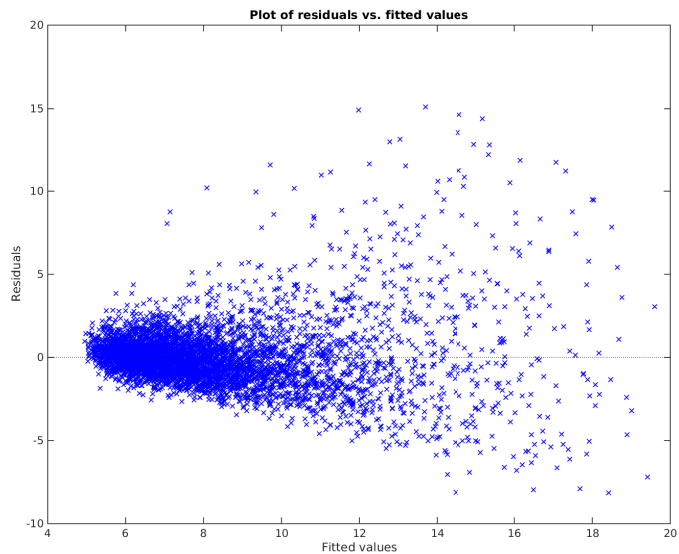}

\caption{Residual plot of PARSEAL.}

\label{Fig: Parseal PWV Model Hete}
\end{figure}

Comparing Figures \ref{Fig: Best Subset PWV Model} and \ref{Fig: Parseal PWV Model},
it is clear that the PARSEAL method is superior to the best subset
selection method. The $R_{adj}^{2}$ of the PARSEAL model is almost
$\%11$ better than the best subset selection method. Both methods
suffer in capturing all the variation and non-linearity in data (compare
Figure \ref{Fig: Best Subset PWV Model BA} to Figure \ref{Fig: Parseal PWV Model BA}
and Figure \ref{Fig: Best Subset PWV Model Hete} to Figure \ref{Fig: Parseal PWV Model Hete}).
However, PARSEAL is better in this respect. The heteroscedasticity
of the best subset selection method is worse than that of the PARSEAL
method (compare Figure \ref{Fig: Best Subset PWV Model Hete} to Figure
\ref{Fig: Parseal PWV Model Hete}). The Bland-Altman limits of agreement
of the PARSEAL method is also better than those of the best subset
selection method (compare Figure \ref{Fig: Best Subset PWV Model BA}
to Figure \ref{Fig: Parseal PWV Model BA}). The latter shows that
the PARSEAL method is a more precise method than the best subset selection
method.

Finally, we again mention that our goal is not to show the best possible
model for PWV, but rather to show that with the same set of inputs,
PARSEAL is superior when compared to other model selection algorithms.

\section{Conclusion and Future Works}

In this paper, we have introduced PARSEAL (Algorithm \ref{Alg: PARSEAL})
by which one can simultaneously capture some of the non-linearities
of the data into the model and also pick the best model. This approach
minimizes the efforts done by an analyst and is virtually automatic.
So far, up to the best of our knowledge, no other algorithm or method
is able to perform these two tasks at the same time automatically.
Furthermore, this method prevents multi-collinearity from entering
the final model.

From the examples presented in this paper, and also the methodology
used by the PARSEAL method, we can claim that our method is one of
the best model selector algorithms. PARSEAL could have versatile applications
in biostatistics as shown by one of the examples in this manuscript. 

In future works, we intend to present some statistical analysis of
the PARSEAL method. In specific, we would like to investigate the
speed of convergence of our method since PARSEAL relies on creating
a large dictionary of inputs. One of the important factors in creating
the mentioned large dictionary of data is the importance limit $\delta$,
Algorithm \ref{Alg: PARSEAL}. The other is the independence limit
$\varsigma$. We would like to quantify optimum values for the importance
limit $\delta$ and also the independence limit $\varsigma$.

We would also like to extend our method to non-integer powers. This
would make PARSEAL capable of better capturing non-integer non-linearities.
Another area of improvement could be adding transformations to the
output variable.

\section{Data Availability}

The data presented, in this paper, can be found at Dryad Digital Repository.
The unique identifier of the data package is doi:10.5061/dryad.c7s7d
.

\section{Competing interests}

\textcolor{black}{We have no competing interests. }

\section{Authors' Contributions}

\textcolor{black}{Peyman Tavallali conceived of the study, carried
out the modeling, programming of the method, carried out the design
of the synthetic examples, and drafted the initial version of the
manuscript; Marianne Razavi helped with the modeling, helped draft
and revise the manuscript; Sean Brady participated in the original
idea of the study, helped draft the manuscript, and revised the manuscript
critically for important intellectual content.}

\textcolor{black}{All authors gave final approval for publication.}

\section{Acknowledgement}

We would like to thank Dr. Niema M. Pahlevan and Prof. Morteza Gharib
for giving us the permission to use the Framingham Heart Study data
in this paper. The Framingham Heart Study is conducted and supported
by the National Heart Lung, and Blood Institute (NHLBI) in collaboration
with Boston University (Contract No. N01- HC-25195). This manuscript
was not prepared in collaboration with investigators of the Framingham
Heart Study and does not necessarily reflect the opinions or conclusions
of the Framingham Heart Study or the NHLBI.

\section{Research Ethics}

The California Institute of Technology and Boston University Medical
Center Institutional Review Boards approved the protocol and all participants
gave written informed consent.

\section{Permission to Carry Out Fieldwork}

This study did not have fieldwork.

\section{Funding}

This work was not funded.

\bibliographystyle{plain}

\section{Figures}

\end{document}